\documentclass[12pt,thmsa]{article}

\input{tcilatex}
\begin{document}

\noindent {\large ACCELERATION WAVES IN THE VON KARMAN PLATE THEORY}%
\footnote{%
Published in: \textit{Integral Methods in Science and Engineering}, Chapman
\& Hall / CRC Research Notes in Mathematics 418, Boca Raton, FL, pp. 131-136
(2000).}\bigskip 

\begin{center}
V. VASSILEV and P. DJONDJOROV

{\small Institute of Mechanics -- Bulgarian Academy of Sciences}

{\small Acad. G. Bontchev St., Block 4, 1113 Sofia, BULGARIA}
\end{center}

\section{Introduction}

The von K\'{a}rm\'{a}n plate theory is governed by two coupled nonlinear
fourth-order partial differential equations in three independent variables
(Cartesian coordinates on the plate middle-plane $x^1,x^2$ and the time $x^3$%
) and two dependent variables (the transversal displacement function $w$ and
Airy's stress function $\mathit{\Phi }$), namely 
\begin{equation}
\begin{tabular}{l}
$D\Delta ^2w-\varepsilon ^{\alpha \mu }\varepsilon ^{\beta \nu }w_{,\alpha
\beta }\mathit{\Phi }_{,\mu \nu }+\rho w_{,33}=0,$ \\ 
$\left( 1/Eh\right) \Delta ^2\mathit{\Phi }+\left( 1/2\right) \varepsilon
^{\alpha \mu }\varepsilon ^{\beta \nu }w_{,\alpha \beta }w_{,\mu \nu }=0,$%
\end{tabular}
\label{1.1}
\end{equation}
where $\Delta $ is the Laplace operator with respect to $x^1$ and $x^2$, $%
D=Eh^3/12(1-\nu ^2)$ is the bending rigidity, $E$ is Young's modulus, $\nu $
is Poisson's ratio, $h$ is the thickness of the plate, $\rho $ is the mass
per unit area of the plate middle-plane, $\delta ^{\alpha \beta }$ is the
Kronecker delta symbol and $\varepsilon ^{\alpha \beta }$ is the alternating
symbol. Here and throughout the work: Greek (Latin) indices range over 1, 2
(1, 2, 3), unless explicitly stated otherwise; the usual summation
convention over a repeated index is used and subscripts after a comma at a
certain function $f$ denote its partial derivatives, that is $%
f_{,i}=\partial f/\partial x^i,\;f_{,ij}=\partial f/\partial x^i\partial x^j$%
, etc.

The von K\'arm\'an equations (\ref{1.1}) describe entirely the motion of a
plate, the membrane stress tensor $N^{\alpha \beta }$, moment tensor $%
M^{\alpha \beta }$, shear-force vector $Q^{\alpha} $, strain tensor $%
E^{\alpha \beta }$ and bending tensor $K_{\alpha \beta }$ being given in
terms of $w$ and $\mathit{\Phi }$ through the following expressions: 
\[
\begin{tabular}{c}
$N^{\alpha \beta }=\varepsilon ^{\alpha \mu }\varepsilon ^{\beta \nu }%
\mathit{\Phi }_{,\mu \nu },\,\;M^{\alpha \beta }=-D\left\{ (1-\nu )\delta
^{\alpha \mu }\delta ^{\beta \nu }+\nu \delta ^{\alpha \beta }\delta ^{\mu
\nu }\right\} w_{,\mu \nu },$ \\ 
$Q^{\alpha} =M_{,\mu }^{\alpha \mu }+N^{\alpha \mu }w_{,\mu },\,\;E^{\alpha
\beta }=(1/Eh)\left\{ (1+\nu )\varepsilon ^{\alpha \mu }\varepsilon ^{\beta
\nu }-v\delta ^{\alpha \beta }\delta ^{\mu \nu }\right\} \mathit{\Phi }%
_{,\mu \nu },\,\;K_{\alpha \beta }=w_{,\alpha \beta }.$%
\end{tabular}
\]

The theory under consideration allows an exact variational formulation, the
von K\'arm\'an equations being the Euler-Lagrange equations [1] associated
with the action functional 
\begin{equation}
I\left[ w,\mathit{\Phi }\right] =\int \int \int Ldx^1dx^2dx^3,\;\,L=T-\Pi
\label{1.2}
\end{equation}
where 
\[
\begin{tabular}{lll}
$\Pi $ & $=$ & $(D/2)\left\{ \left( \Delta w\right) ^2-\left( 1-\nu \right)
\varepsilon ^{\alpha \mu }\varepsilon ^{\beta \nu }w_{,\alpha \beta }w_{,\mu
\nu }\right\} $ \\ 
& $-$ & $(1/2Eh)\left\{ \left( \Delta \mathit{\Phi }\right) ^2-\left( 1+\nu
\right) \varepsilon ^{\alpha \mu }\varepsilon ^{\beta \nu }\mathit{\Phi }%
_{,\alpha \beta }\mathit{\Phi }_{,\mu \nu }\right\} +(1/2)\varepsilon
^{\alpha \mu }\varepsilon ^{\beta \nu }\mathit{\Phi }_{,\alpha \beta
}w_{,\mu }w_{,\nu }, $%
\end{tabular}
\label{1.3} 
\]
is the strain energy per unit area of the plate middle-plane and 
\[
T=\left( \rho /2\right) \left( w_{,3}\right) ^2,\label{1.4} 
\]
is the kinetic energy per unit area of the plate middle-plane.

\section{Conservation laws}

In the recent paper [2], all Lie point symmetries of system (\ref{1.1}) are
shown to be variational symmetries of the functional (\ref{1.2}), and all
corresponding (via Noether's theorem) conservation laws admitted by the
smooth solutions of the von K\'arm\'an equations are established. Each such
conservation law is a linear combination of the basic linearly independent
conservation laws 
\[
\frac{\partial \mathit{\Psi }_{(j)}}{\partial x^3}+\frac{\partial
P_{(j)}^{\mu} }{\partial x^{\mu} }=0\;\,\left( j=1,2,\ldots ,14\right) , 
\]
whose densities $\mathit{\Psi }_{(j)}$ and fluxes $P_{(j)}^{\mu} $ are
presented (together with the generators of the respective symmetries) on the
Table 1 below in terms of $Q^{\alpha} ,\,M^{\alpha \beta },\,G^{\alpha \beta
}$ and $F^{\alpha} $, 
\[
G^{\alpha \beta }=\left( 1/Eh\right) \left\{ (1+\nu )\delta ^{\alpha \mu
}\delta ^{\beta \nu }-\nu \delta ^{\alpha \beta }\delta ^{\mu \nu }\right\} 
\mathit{\Phi }_{,\mu \nu }-\left( 1/2\right) \varepsilon ^{\alpha \mu
}\varepsilon ^{\beta \nu }w_{,\mu }w_{,\nu },\quad F^{\alpha} =G_{,\nu
}^{\alpha \nu }. 
\]

\noindent 
\begin{tabular}[t]{l}
Table 1 Conservation laws
\end{tabular}

\noindent 
\begin{tabular}{|p{4cm}p{10.6cm}|}
\hline
\multicolumn{1}{|l}{$w$ - translations} & \multicolumn{1}{r|}{\textbf{%
transversal linear momentum }(first von K\'{a}rm\'{a}n eqn)} \\ 
$X_1=\frac \partial {\partial \,w}$ & 
\begin{tabular}{l}
$P_{(1)}^{\alpha} =-Q^{\alpha} ,\;\mathit{\Psi }_{(1)}=\rho \,w_{,3}$%
\end{tabular}
\\ \hline
\end{tabular}

\noindent 
\begin{tabular}{|p{4cm}p{10.6cm}|}
\hline
\multicolumn{1}{|l}{$\mathit{\Phi }$ - translations} & \multicolumn{1}{r|}{%
\textbf{compatibility condition }(second von K\'{a}rm\'{a}n eqn)} \\ 
$X_{14}=\frac \partial {\partial \mathit{\Phi }}$ & 
\begin{tabular}{l}
$P_{(14)}^{\alpha} =F^{\alpha} ,\;\mathit{\Psi }_{(14)}=0$%
\end{tabular}
\\ \hline
\end{tabular}

\noindent $
\begin{tabular}{|p{4cm}p{10.6cm}|}
\hline
\multicolumn{1}{|l}{time - translations} & \multicolumn{1}{r|}{\textbf{energy%
}} \\ 
$X_4=\frac \partial {\partial \,x^3}$ & 
\begin{tabular}{l}
$P_{(4)}^{\alpha} =-w_{,3}Q^{\alpha} -\mathit{\Phi }_{,3}F^{\alpha}
+w_{,3\beta }M^{\alpha \beta }+\mathit{\Phi }_{,3\beta }G^{\alpha \beta }$
\\ 
$\mathit{\Psi }_{(4)}=T+\Pi $%
\end{tabular}
\\ \hline
\end{tabular}
$

\noindent $
\begin{tabular}{|p{4cm}p{10.6cm}|}
\hline
\multicolumn{1}{|l}{$x^1\,\&\,x^2$- translations} & \multicolumn{1}{r|}{%
\textbf{wave momentum}} \\ 
$X_2=\frac \partial {\partial \,x^1}$ & 
\begin{tabular}{l}
$P_{(2)}^{\alpha} =\delta ^{\alpha 1}L+w_{,1}Q^{\alpha} +\mathit{\Phi }%
_{,1}F^{\alpha} -w_{,1\beta }M^{\alpha \beta }-\mathit{\Phi }_{,1\beta
}G^{\alpha \beta }$ \\ 
$\mathit{\Psi }_{(2)}=-\rho \,w_{,1}w_{,3}$%
\end{tabular}
\\ 
\multicolumn{1}{|l}{$X_3=\frac \partial {\partial \,x^2}$} & 
\multicolumn{1}{l|}{
\begin{tabular}{l}
$P_{(3)}^{\alpha} =\delta ^{\alpha 2}L+w_{,2}Q^{\alpha} +\mathit{\Phi }%
_{,2}F^{\alpha} -w_{,2\beta }M^{\alpha \beta }-\mathit{\Phi }_{,2\beta
}G^{\alpha \beta }$ \\ 
$\mathit{\Psi }_{(3)}=-\rho \,w_{,2}w_{,3}$%
\end{tabular}
} \\ \hline
\end{tabular}
$

\noindent $
\begin{tabular}{|p{4cm}p{10.6cm}|}
\hline
\multicolumn{1}{|l}{rotations} & \multicolumn{1}{r|}{\textbf{moment of the
wave momentum}} \\ 
$X_6=x^2\frac \partial {\partial \,x^1}-x^1\frac \partial {\partial \,x^2}$
& 
\begin{tabular}{l}
$P_{(6)}^{\alpha} =x^2P_{(2)}^{\alpha} -x^1P_{(3)}^{\alpha} +\varepsilon
_{\nu} ^{\;\mu }w_{,\mu }M^{\alpha \nu }+\varepsilon _{\nu }^{\;\mu }\mathit{%
\Phi } _{,\mu }G^{\alpha \nu }$ \\ 
$\mathit{\Psi }_{(6)}=x^2\mathit{\Psi }_{(2)}-x^1\mathit{\Psi }_{(3)}$%
\end{tabular}
\\ \hline
\end{tabular}
$

\noindent $
\begin{tabular}{|p{4cm}p{10.6cm}|}
\hline
\multicolumn{1}{|l}{rigid body rotations} & \multicolumn{1}{r|}{\textbf{%
angular momentum}} \\ 
$X_7=x^1\frac \partial {\partial \,w}$ & 
\begin{tabular}{c}
$P_{(7)}^{\alpha} =M^{\alpha 1}-x^1Q^{\alpha} +w\varepsilon ^{\alpha \nu }%
\mathit{\Phi }_{,\nu 2},\;\mathit{\Psi }_{(7)}=\rho x^1w_{,3}$%
\end{tabular}
\\ 
\multicolumn{1}{|l}{$X_8=x^2\frac \partial {\partial w}$} & 
\multicolumn{1}{l|}{
\begin{tabular}{l}
$P_{(8)}^{\alpha} =M^{\alpha 2}-x^2Q^{\alpha} +w\varepsilon ^{\nu \alpha }%
\mathit{\Phi }_{,\nu 1},\;\mathit{\Psi }_{(8)}=\rho x^2w_{,3}$%
\end{tabular}
} \\ \hline
\end{tabular}
$

\noindent $
\begin{tabular}{|p{4cm}p{10.6cm}|}
\hline
\multicolumn{1}{|l}{scaling} & \multicolumn{1}{l|}{} \\ 
$X_5=x^{\mu} \frac \partial {\partial \,x^{\mu} }+2x^3\frac \partial
{\partial \,x^3}$ & 
\begin{tabular}{l}
$P_{(5)}^{\alpha} =x^1P_{(2)}^{\alpha} +x^2P_{(3)}^{\alpha}
-2x^3P_{(4)}^{\alpha} -w_{,\beta }M^{\alpha \beta }-\mathit{\Phi }_{,\beta
}G^{\alpha \beta }$ \\ 
$\mathit{\Psi }_{(5)}=x^1\mathit{\Psi }_{(2)}+x^2\mathit{\Psi }_{(3)}-2x^3%
\mathit{\Psi }_{(4)}$%
\end{tabular}
\\ \hline
\end{tabular}
$

\noindent $
\begin{tabular}{|p{4cm}p{10.6cm}|}
\hline
\multicolumn{1}{|l}{Galilean boost} & \multicolumn{1}{r|}{\textbf{%
center-of-mass theorem}} \\ 
$X_9=x^3\frac \partial {\partial w}$ & 
\begin{tabular}{l}
$P_{(9)}^{\alpha} =-x^3Q^{\alpha} ,\;\mathit{\Psi }_{(9)}=\rho
\,(x^3w_{,3}-w)$%
\end{tabular}
\\ \hline
\end{tabular}
$

\noindent 
\begin{tabular}{|p{4cm}p{10.6cm}|}
\hline
$X_{10}=x^1x^3\frac \partial {\partial w}$ & 
\begin{tabular}{l}
$P_{(10)}^{\alpha} =x^3P_{(7)}^{\alpha} ,\;\mathit{\Psi }_{(10)}=x^1\mathit{%
\Psi }_{(9)}$%
\end{tabular}
\\ 
\multicolumn{1}{|l}{$X_{11}=x^2x^3\frac \partial {\partial w}$} & 
\multicolumn{1}{l|}{
\begin{tabular}{l}
$P_{(11)}^{\alpha} =x^3P_{(8)}^{\alpha} ,\;\mathit{\Psi }_{(11)}=x^2\mathit{%
\Psi }_{(9)}$%
\end{tabular}
} \\ 
\multicolumn{1}{|l}{$X_{12}=x^1\frac \partial {\partial \mathit{\Phi }}$} & 
\multicolumn{1}{l|}{
\begin{tabular}{l}
$P_{(12)}^{\alpha} =x^1F^{\alpha} -G^{\alpha 1},\;\mathit{\Psi }_{(12)}=0$%
\end{tabular}
} \\ 
\multicolumn{1}{|l}{$X_{13}=x^2\frac \partial {\partial \mathit{\Phi }}$} & 
\multicolumn{1}{l|}{
\begin{tabular}{l}
$P_{(13)}^{\alpha} =x^2F^{\alpha} -G^{\alpha 2},\;\mathit{\Psi }_{(13)}=0$%
\end{tabular}
} \\ \hline
\end{tabular}

\section{Balance laws}

Given a region $\Omega $ in the plate middle-plane with sufficiently smooth
boundary $\Sigma $ of outward unit normal $n_{\alpha} $, a balance law 
\begin{equation}
\frac d{dt}\int\limits_\Omega \mathit{\Psi }_{(j)}dx^1dx^2+\int\limits_%
\Sigma P_{(j)}^{\alpha} n_{\alpha} d\Sigma =0,  \label{2.4}
\end{equation}
corresponds to each of the conservation laws listed in Table 1. It holds,
just as the respective conservation law, for every smooth solution of the
von K\'{a}rm\'{a}n equations.

The balance laws are applicable even if $\Omega $ is intersected by a
discontinuity (singular) manifold (on which the corresponding densities $%
\mathit{\Psi }_{(j)}$ and fluxes $P_{(j)}^{\alpha} $ may suffer jump
discontinuities) provided that the integrals exist. We are ready now to
extend the ``continuous`` von K\'arm\'an plate theory so as to cover
situations when some physical quantities suffer jump discontinuities at a
certain curve.

\section{Acceleration waves}

\begin{definition}
A discontinuity solution of the von K\'{a}rm\'{a}n equations is a couple of
functions $(w,\mathit{\Phi })$, defined in a certain region $\Omega $, such
that the two balance laws corresponding to the von K\'{a}rm\'{a}n equations
themselves, namely 
\begin{equation}
\frac d{dt}\int\limits_{{\tilde{\Omega}}}\rho w_{,3}dx^1dx^2-\int\limits_{%
\tilde{\Sigma}}Q^{\alpha} \tilde{n}_{\alpha} d\Sigma =0,\,\;\int\limits_{%
\tilde{\Sigma}}F^{\alpha} \tilde{n}_{\alpha} d\Sigma =0,  \label{3.1}
\end{equation}
hold $\forall \,{\tilde{\Omega}}\subset \Omega $ with boundary $\tilde{\Sigma%
}$ of outward unit normal $\tilde{n}_{\alpha} $, and $(w,\mathit{\Phi })$ is
a solution of the (local) von K\'{a}rm\'{a}n equations (\ref{1.1}) almost
everywhere in $\Omega $ except for a moving curve $\Gamma $ at which some of
the derivatives of $w$ or $\mathit{\Phi }$ have jumps.
\end{definition}

\begin{definition}
A discontinuity solution of the von K\'{a}rm\'{a}n equations is an
acceleration wave if at the wave front -- a smoothly propagating connected
singular curve $\Gamma $ -- 
\[
\Gamma :\gamma (x^1,x^2,x^3)=0,\;\,(x^1,x^2)\in \Omega \subset \mathbf{R}%
^2,\;\,x^3\in \mathbf{R}^{+},\;\,\gamma \in C^1(\Omega \times \mathbf{R}%
^{+}),\label{3.2} 
\]
we have 
\begin{equation}
\lbrack w]=[\mathit{\Phi }]=[w_{,i}]=[\mathit{\Phi }_{,i}]=0,\;\,[w_{,33}]%
\neq 0.  \label{3.3}
\end{equation}
(Here and in what follows, the square brackets are used to denote the jump
of any field $f$ across the curve $\Gamma $, i.e., $[f]=f_2-f_1,$ where $f_2$
and $f_1$ are the limit values of $f$ behind $\Gamma $ and ahead of $\Gamma $%
.)
\end{definition}

The moving curve $\Gamma $ divides the region $\Omega $ into two parts $%
\Omega ^{+}$ and $\Omega ^{-}$ and forms the common border between them. It
is assumed that ahead of the wave front (in the region $\Omega ^{+}$) we
have the known unperturbed fields $w^{+}(x^1,x^2,x^3)$, $\mathit{\Phi }%
^{+}(x^1,x^2,x^3)$ and behind it (in the region $\Omega ^{-}$) -- the
unknown perturbed fields $w^{-}(x^1,x^2,x^3)$, $\mathit{\Phi }%
^{-}(x^1,x^2,x^3)$. At the wave front $\Gamma $, we have the jump conditions
(\ref{3.3}).

The jumps of the derivatives of $w$ and $\mathit{\Phi }$ across $\Gamma $
are permissible if they obey the compatibility conditions following by
Hadamard's lemma [3]. Thus

\begin{proposition}
If $[w_{,33}]\neq 0$, then 
\[
\lbrack w_{,\alpha \beta }]=\lambda n_{\alpha} n_{\beta} ,\;\,[w_{,\alpha
3}]=-\lambda Cn_{\alpha} ,\;\,\,[w_{,33}]=\lambda C^2,\label{3.4} 
\]
where $\lambda $ is an arbitrary factor, $C$ and $n_{\alpha} $, 
\[
C=-\left| \nabla \gamma \right| ^{-1}\partial \gamma /\partial
x^3,\;\,n_{\alpha} ={\left| \nabla \gamma \right| ^{-1}\partial \gamma /}%
\partial x^{\alpha} ,\;\,\left| \nabla \gamma \right| =\sqrt{(\partial
\gamma /\partial x^1)^2+({\partial }\gamma /\partial x^2)^2},\label{3.5} 
\]
are the speed of displacement and the direction of propagation of the wave
front $\Gamma $.
\end{proposition}

\begin{proposition}
If at least one of the third derivatives of $w$ suffers a jump at $\Gamma $,
then the compatibility conditions for the jumps of the third derivatives of
the displacement field across $\Gamma $ are: 
\[
\begin{tabular}{lll}
$\left[ w_{,\alpha \beta \gamma }\right] $ & $=$ & $\lambda ^{*}n_{\alpha}
n_{\beta} n_{\gamma} +{\partial \lambda /\partial s}\left( n_{\alpha}
n_{\beta} t_{\gamma} +n_{\alpha} t_{\beta} n_{\gamma} +t_{\alpha} n_{\beta}
n_{\gamma} \right) $ \\ 
& $+$ & $\lambda a\left( t_{\alpha} t_{\beta} n_{\gamma} +t_{\alpha}
n_{\beta} t_{\gamma} +n_{\alpha} t_{\beta} t_{\gamma} \right) ,$%
\end{tabular}
\label{3.6} 
\]
where $t_{\alpha} $ is the unit tangent vector to $\Gamma $, $\lambda ^{*}$
is an arbitrary factor, while $\lambda =\left[ w_{,\alpha \beta }\right]
n^{\alpha} n^{\beta} $ and $a=t_{\alpha} \partial n^{\alpha} /\partial s$, $s
$ being the natural parameter (arc-length) of the curve $\Gamma $.
\end{proposition}

\begin{proposition}
If at least one of the second derivatives of $\mathit{\Phi }$ suffers a jump
at $\Gamma $, then 
\[
\lbrack \mathit{\Phi }_{,\alpha \beta }]=\mu n_{\alpha} n_{\beta} ,\quad [%
\mathit{\Phi }_{,\alpha 3}]=-\mu Cn_{\alpha} ,\quad [\mathit{\Phi }%
_{,33}]=\mu C^2,\label{3.7} 
\]
where $\mu $ is an arbitrary factor, are the compatibility conditions for
the jumps of the second derivatives of the stress field across $\Gamma $.
\end{proposition}

\begin{proposition}
If at least one of the third derivatives of $\mathit{\Phi }$ suffer a jump
at $\Gamma $, then the compatibility conditions for the jumps of the third
derivatives of the stress field across $\Gamma $ are: 
\[
\begin{tabular}{lll}
$\left[ \mathit{\Phi }_{,\alpha \beta \gamma }\right] $ & $=$ & $\mu
^{*}n_{\alpha} n_{\beta} n_{\gamma} +{\partial \mu /\partial s}\left(
n_{\alpha} n_{\beta} t_{\gamma} +n_{\alpha} t_{\beta} n_{\gamma} +t_{\alpha}
n_{\beta} n_{\gamma} \right) $ \\ 
& $+$ & $\mu a\left( t_{\alpha} t_{\beta} n_{\gamma} +t_{\alpha} n_{\beta}
t_{\gamma}+n_{\alpha} t_{\beta} t_{\gamma} \right) ,$%
\end{tabular}
\label{3.8} 
\]
where $\mu =[\mathit{\Phi }_{,\alpha \beta }]n^{\alpha} n^{\beta} $ and $\mu
^{*} $ is an arbitrary factor.
\end{proposition}

According to the divergence theorem (see e.g. [4]), a couple of functions $%
(w,\,\mathit{\Phi })$ suffering jump discontinuities at a singular curve $%
\Gamma $ is a discontinuity solution of the von K\'arm\'an equations in the
sense of Definition 1 iff the following jump conditions 
\begin{equation}
C\left[ \rho w_{,3}\right] +\left[ Q^{\alpha} \right] n_{\alpha} =0,\quad
\left[ F^{\alpha} \right] n_{\alpha} =0,  \label{3.9}
\end{equation}
hold at $\Gamma $, and a balance law of form (\ref{2.4}) holds on this
solution iff at $\Gamma $: 
\begin{equation}
C\left[ \mathit{\Psi }_{\left( j\right) }\right] -\left[ P_{\left( j\right)
}^{\alpha} \right] n_{\alpha} =0.  \label{3.10}
\end{equation}

Definition 2, Propositions 1, 2, 3, 4 and jump conditions (\ref{3.9}) imply
that:

\begin{proposition}
If an acceleration wave in the von K\'{a}rm\'{a}n plate theory is such that $%
[w_{,\alpha \beta \gamma }]\neq 0$ ($[\mathit{\Phi }_{,\alpha \beta \gamma
}]\neq 0$) at the curve of discontinuity $\Gamma $, then $\lambda
^{*}=-\lambda a$ ($\mu ^{*}=-\mu a$).
\end{proposition}

Given a discontinuity solution of the von K\'{a}rm\'{a}n equations, the two
corresponding balance laws (\ref{3.1}) being satisfied, the other balance
laws do not necessarily hold for this solution. The jump conditions
associated with the most important conservation laws from Table 1 are
derived using (\ref{3.10}) and presented on the Table 2 below, where $%
X_j^{+}(w_{,\beta })$ and $X_j^{+}(\mathit{\Phi }_{,\beta })$ denote the
limit values of $X_j(w_{,\beta })$ and $X_j(\mathit{\Phi }_{,\beta })$ {%
ahead of} $\Gamma $:

\noindent 
\begin{tabular}[t]{l}
Table 2 Jump conditions
\end{tabular}

\noindent $
\begin{tabular}{|p{4cm}p{10.6cm}|}
\hline
\multicolumn{1}{|l}{time - translations} & \multicolumn{1}{r|}{\textbf{energy%
}} \\ 
$X_4=\frac \partial {\partial \,x^3}$ & 
\begin{tabular}{l}
$\frac C2\left( D\lambda ^2-\frac{\mu ^2}{Eh}\right) =\left( D\lambda
X_4^{+}(w_{,\beta })-\frac \mu {Eh}X_4^{+}(\mathit{\Phi }_{,\beta })\right)
n^{\beta} $%
\end{tabular}
\\ \hline
\end{tabular}
$

\noindent $
\begin{tabular}{|p{4cm}p{10.6cm}|}
\hline
\multicolumn{1}{|l}{$x^1$\thinspace \&\thinspace $x^2$- translations} & 
\multicolumn{1}{r|}{\textbf{wave momentum}} \\ 
$X_2=\frac \partial {\partial \,x^1}$ & $
\begin{tabular}{l}
$\frac 12\left( D\lambda ^2-\frac{\mu ^2}{Eh}\right) n^1=-\left( D\lambda
X_2^{+}(w_{,\beta })-\frac \mu {Eh}X_2^{+}(\mathit{\Phi }_{,\beta })\right)
n^{\beta} $%
\end{tabular}
$ \\ 
\multicolumn{1}{|l}{$X_3=\frac \partial {\partial \,x^2}$} & 
\multicolumn{1}{l|}{$
\begin{tabular}{l}
$\frac 12\left( D\lambda ^2-\frac{\mu ^2}{Eh}\right) n^2=-\left( D\lambda
X_3^{+}(w_{,\beta })-\frac \mu {Eh}X_3^{+}(\mathit{\Phi }_{,\beta })\right)
n^{\beta} $%
\end{tabular}
$} \\ \hline
\end{tabular}
$

\noindent $
\begin{tabular}{|p{4cm}p{10.6cm}|}
\hline
\multicolumn{1}{|l}{rotations} & \multicolumn{1}{r|}{\textbf{moment of the
wave momentum}} \\ 
$X_6=x^2\frac \partial {\partial \,x^1}-x^1\frac \partial {\partial \,x^2}$
& 
\begin{tabular}{l}
$\frac 12\varepsilon _{\;\beta }^{\alpha} n_{\alpha} x^{\beta} \left(
D\lambda \,^2-\frac{\mu \,^2}{Eh}\right) =$ \\ 
$-\left\{ D\lambda \left( \varepsilon _{\;\beta }^{\alpha} w_{,\alpha
}+X_6^{+}(w_{,\beta })\right) -\frac \mu {Eh}\left( \varepsilon _{\;\beta
}^{\alpha} \mathit{\Phi }_{,\alpha }+X_6^{+}(\mathit{\Phi }_{,\beta
})\right) \right\} n^{\beta} $%
\end{tabular}
\\ \hline
\end{tabular}
$

\noindent $
\begin{tabular}{|p{4cm}p{10.6cm}|}
\hline
\multicolumn{1}{|l}{scaling} & \multicolumn{1}{l|}{} \\ 
$X_5=x^{\mu} \frac \partial {\partial \,x^{\mu} }+2x^3\frac \partial
{\partial \,x^3}$ & 
\begin{tabular}{l}
$\frac 12\left( x^{\alpha} n_{\alpha} -2Cx^3\right) \left( D\lambda \,^2-%
\frac{\mu \,^2}{Eh}\right) =$ \\ 
$-\left\{ D\lambda \left( w_{,\beta }+X_5^{+}(w_{,\beta })\right) -\frac \mu
{Eh}\left( \mathit{\Phi }_{,\beta }+X_5^{+}(\mathit{\Phi }_{,\beta })\right)
\right\} n^{\beta} $%
\end{tabular}
\\ \hline
\end{tabular}
$

\begin{tabular}{l}
\end{tabular}

\noindent The center-of-mass theorem holds for any discontinuity solutions
of the von K\'{a}rm\'{a}n equations. The balance laws, associated with the
infinitesimal symmetries $X_7$, $X_8$, $X_{10}$ and $X_{11}$ hold iff $%
\lambda =0$, while those associated with $X_{12}$ and $X_{13}$ -- iff $\mu
=0 $. For this reason, there do not exist acceleration waves in the von
K\'{a}rm\'{a}n plate theory satisfying all balance laws.

Obviously, when dealing with discontinuity solutions, from physical point of
view it seems reasonably that at least the balance of energy should hold in
addition to the balance laws corresponding to the fundamental equations
considered. Observing Table 2 it is evident that for acceleration waves
propagating into an undisturbed plate the balance of energy implies also the
balances of wave momentum, moment of the wave momentum as well as the
balance related to the scaling symmetry.

\section{Examples}

As an example, we consider acceleration waves such that behind and ahead of
the wave front the plate motion is described by solutions of the von
K\'arm\'an equations invariant under the group generated by $X_3$ and $%
X_2+(1/c)X_4$, where $c$ is an arbitrary constant. The most general form of
such group-invariant solutions is 
\[
\begin{tabular}{l}
$w=u(\xi )=u_0+u_1\xi +u_2\sin \omega \xi +u_3\cos \omega \xi ,$ \\ 
$\mathit{\Phi }=\varphi (\xi )=\varphi _0+\varphi _1\xi +\varphi _2\xi
^2+\varphi _3\xi ^3,$%
\end{tabular}
\label{InvSol} 
\]
where $\xi =x^1-cx^3$, $u_j$, $\varphi _j$ are arbitrary constants, and $%
\omega =c\sqrt{\rho /D}$. Let ($u^{+}$,$\varphi ^{+}$) -- an arbitrary
solution of that kind -- describes the plate motion ahead of the wave front.
Then, Definition 2 and Propositions 1 to 4 imply that each acceleration wave
of the type considered reads 
\begin{equation}
u=\left\{ 
\begin{tabular}{ll}
$u^{+}+c_1(1-\cos \omega \xi ),$ & $\;\xi <0,$ \\ 
$u^{+},$ & $\;\xi >0,$%
\end{tabular}
\right. \qquad \varphi =\left\{ 
\begin{tabular}{ll}
$\varphi ^{+}+c_2\xi ^2,$ & $\;\xi <0,$ \\ 
$\varphi ^{+},$ & $\;\xi >0,$%
\end{tabular}
\right.  \label{Accwave}
\end{equation}
where $c_1$ and $c_2$ are arbitrary constants, but $c_1\neq 0$; the wave
front in this case is the moving straight line $\Gamma :\xi =0$. In general,
however, an acceleration wave of form (\ref{Accwave}) does not satisfy the
balance laws other than (\ref{3.1}). Indeed, after a little manipulation,
the jump conditions from Table 2 simplify to 
\begin{eqnarray}
DEh\omega ^4c_1(c_1-2u_3^{+}) &=&4c_2(c_2+2\varphi _2^{+}),  \label{EJ} \\
DEh\omega ^2c_1(u_1^{+}+\omega u_2^{+}) &=&2c_2\varphi _1^{+},  \label{OJ}
\end{eqnarray}
where $u_j^{+}$, $\varphi _j^{+}$ are the constants in $u^{+}$ and $\varphi
^{+}$, respectively. The jump condition (\ref{EJ}) is necessary and
sufficient for the balances of energy, wave momentum and moment of wave
momentum to hold, while the balance related to $X_5$ requires both (\ref{EJ}%
) and (\ref{OJ}). The second relation is treated in a different manner
according to the acceleration wave under consideration. If the wave is such
that $c_2=0$, then (\ref{OJ}) holds for this wave only if $u_1^{+}=-\omega
u_2^{+}$. On the other hand, if we consider waves with $c_2\neq 0$, then
choosing the coefficient $\varphi _1^{+}$ in a suitable manner, we could
satisfy (\ref{OJ}) identically (note that adding a linear function of the
independent variables to Airy's stress function does not change the membrane
stress tensor $N^{\alpha \beta }$). Hence, the balance associated with the
scaling symmetry $(X_5)$ holds only for acceleration waves of form (\ref
{Accwave}) satisfying (\ref{EJ}) and which are such that either $c_2\neq 0$
or $u_1^{+}=-\omega u_2^{+}$.

Another example, discussed in details in [2], is an axisymmetrically
expanding acceleration wave composed by solutions of the von K\'{a}rm\'{a}n
equations that are joined invariants of the rotation ($X_6$) and scaling ($%
X_5$) symmetries.\smallskip 

\noindent \textit{Acknowledgments}. This work was supported by Contract No.
MM 517/1995 with NSF, Bulgaria.

\end{document}